\documentclass[10pt]{article}

\usepackage{amsmath}
\usepackage{amssymb}

\usepackage{graphicx}

\usepackage{cite}

\usepackage{color} 

\usepackage[labelfont=bf,labelsep=period,justification=raggedright]{caption}

\makeatletter
\renewcommand{\@biblabel}[1]{\quad#1.}
\makeatother

\date{}

\begin{document}

% Title must be 150 characters or less
\begin{flushleft}
{\Large
%\textbf{Social unrest in Spain: analyzing the time-stamped network of political communication}
\textbf{Structural and Dynamical Patterns on Online Social Networks: the Spanish May 15th Movement as a case study}
}
% Insert Author names, affiliations and corresponding author email.
\\
Javier Borge-Holthoefer$^{1}$, 
Alejandro Rivero$^{1}$,
I\~nigo Garc\'ia$^{3}$,
Elisa Cauh\'e$^{1}$,
Alfredo Ferrer$^{1}$,
Dar\'io Ferrer$^{1}$,
David Francos$^{1}$,
David I\~niguez$^{1,4}$,
Mar\'ia Pilar P\'erez$^{1}$,
Gonzalo Ruiz$^{1}$,
Francisco Sanz$^{1}$,
Ferm\'in Serrano$^{1}$,
Cristina Vi\~nas$^{1}$,
Alfonso Taranc\'on$^{1,2}$,
Yamir Moreno$^{1,2,5,\ast}$
\\
\bf{1} Instituto de Biocomputaci\'on y F\'\i sica de Sistemas 
Complejos (BIFI), Universidad de Zaragoza, Mariano Esquillor s/n, 50018 Zaragoza, Spain
\\
\bf{2} Departamento de F\'{\i}sica Te\'orica, Universidad de Zaragoza, 50009 Zaragoza, Spain
\\
\bf{3} Cierzo Development S.L., Zaragoza, Spain
\\
\bf{4} Fundaci\'on ARAID, Diputaci\'on General de Arag\'on, Zaragoza, Spain
\\
\bf{5} Complex Networks and Systems Lagrange Lab, Institute for Scientific Interchange, Viale S. Severo 65, 10133 Torino, Italy
\\
$\ast$ E-mail: yamir.moreno@gmail.com
\end{flushleft}

% Please keep the abstract between 250 and 300 words
\section*{Abstract}

The number of people using online social networks in their everyday life is continuously growing at a pace never saw before. This new kind of communication has an enormous impact on opinions, cultural trends, information spreading and even in the commercial success of new products. More importantly, social online networks have revealed as a fundamental organizing mechanism in recent country-wide social movements. In this paper, we provide a quantitative analysis of the structural and dynamical patterns emerging from the activity of an online social network around the ongoing May 15th (15M) movement in Spain. Our network is made up by users that exchanged tweets in a time period of one month, which includes the birth and stabilization of the 15M movement. We characterize in depth the growth of such dynamical network and find that it is scale-free with communities at the mesoscale. We also find that its dynamics exhibits typical features of critical systems such as robustness and power-law distributions for several quantities. Remarkably, we report that the patterns characterizing the spreading dynamics are asymmetric, giving rise to a clear distinction between information sources and sinks. Our study represent a first step towards the use of data from online social media to comprehend modern societal dynamics.

% Please keep the Author Summary between 150 and 200 words
% Use first person. PLoS ONE authors please skip this step. 
% Author Summary not valid for PLoS ONE submissions.   
%\section*{Author Summary}

\section*{Introduction}

Modern online socio-technological systems are producing a deep change in our traditional networking paradigms and in the way we communicate with each other. At the same time, online social media constitute nowadays efficient and fast means to group together many social agents around a common issue. In this way, new types of economic, financial and social phenomena are arising. An example of the latter is given by the so-called Arab revolts, which have materialized thanks to these new communication platforms. The protests have been mediated by the use of social networks such as Facebook, Twitter and YouTube, which have been critical for the birth and consolidation of campaigns involving strikes, demonstrations, marches and rallies.

On the other hand, online social networks not only modify in a radical way the dynamics of information and opinion spreading, but are also making our world even more global. More importantly, these platforms generate an enormous amount of time-stamped data, making it possible for the first time to study the fast dynamics associated to different spreading processes at a system-wide scale. These novel and rich data niches allow testing different social dynamics and models that would otherwise be highly elusive with traditional data-gathering methods. Additionally, the availability of data enables the study of phenomena that take place on time scales ranging from a few minutes or hours to a year-long duration. An example of the former kind of fast dynamics is given by large sport or cultural events, whereas cooperative content production such as the case of Wikipedia typically occurs in months or even years, thus, in a much slower time scale.

In this paper, we study the structural and dynamical patterns of the network made up by twitter users who have been involved in a social phenomenon that is currently taking place in Spain: the so-called May 15th movement (henceforth referred to as 15M). This movement-in-the-making had been brewing for a while in the social media, but took off on May 15th when the first demonstrators camped in a central square in Madrid, Spain. From that moment on, the protests and camps spread throughout the country. As many of the adherents are online social media users, the growth and stabilization of the movement was closely reflected in time-stamped data of twitter messages, which we have gathered and analyzed. This will allow us to elucidate the mechanisms driving the emergence of this kind of social phenomenon, and whether it shares dynamical and structural features with other natural, social and technological processes \cite{boccaletti06,dgm08}. Additionally, on more general scientific grounds, a social phenomenon like the 15M movement is an excellent opportunity to understand network formation processes and online spreading dynamics. The ultimate aim is to further advance our understanding of this kind of dynamics and eventually be able to make predictions based on real time data monitoring.

In what follows, we present the results of our analysis. On the one hand, we statistically characterize the structural patterns of the network of users who sent or received tweets containing keywords related to the 15M movement. We find that this network displays the typical features of other networks in Nature such as scale-free degree distributions, a community structure at the mesoscale and high structural robustness \cite{boccaletti06,dgm08}. On the other hand, we have also analyzed the dynamical patterns characterizing the spreading of information over the 15M network.  Our results show that the 15M diffusion dynamics is highly asymmetric. Admittedly, a relative large fraction of the system is actively trafficking, but a great part of the overall traffic is delivered to a few users that do not pass them anymore, thus constituting a sort of information sinks. We round off our analysis by comparing our results with those reported in the literature for other kinds of online dynamical processes.

\section*{Methods}

The data used in this study is a set of messages (tweets) that were publicly exchanged through {\em www. twitter.com}. The whole time-stamped data collected comprises a period of one month (between April 25th, 2011 at 00:03:26 and May 26th, 2011 at 23:59:55) and it was archived by a local start-up company, {\em Cierzo Development Ltd} using the SMMART Platform. This platform is evolving into a new concept called "Open Social CRM," which combines concepts in monitoring tools, CRM tools, social tools and a philosophy of open innovation. The company restricts its collection to messages in Spanish language that come preferentially from users within or related to Spain. The internals of data collection are private to the company, but basically 23 hours of data are homogeneously collected each day, always leaving the same timeframe (16:00 to 17:00 CET time) to readjust the database due to the introduction of new Spanish nodes, purging of the non-Spanish related ones, etc. 

To filter out the whole sample and choose only those messages related to the 15M movement, we selected 70 keywords ({\em hashtags}, see {\em Supplementary Information}) which were systematically used by the adherents to the demonstrations and camps. Next, the extracted sample was examined for missing hashtags $-$ of the top ten only one of them was not in the selected set, this being likely related to its bilingual nature {\em $\#$acampadabcn}. The filtered data set appears to be representative enough of the total traffic related to the 15M movement produced during the period analyzed. As a matter of fact, a comparison with other databases, such as {\em topsy.com}, which aims to collect the whole set of twitter messages, shows that for the same period there were about 390.000 messages with the word "{\em acampadasol}" and 190.000 for the hashtag "{\em $\#$nolesvotes}". Our sample is made up of 189.000 and 66.000 messages and hashtags, respectively, i.e., roughly above a third of the total number of messages. 

Once this process is finished, the final sample consists of 581.749 tweets, out of which 46.557 were identified as {\em retweets} of unknown origin, and therefore were discarded. On its turn, these tweets were produced by 85.851 different users. To complete the data set, we located the references to other users inside each message. These references are marked in the system by an arobase, "$@$username". A user receives a notification, usually via email, each time a mention happens, and the messages having mentions are also copied to a special tab in the user interface. The total number of messages having at least a reference was 151.222. In some cases, the tweet is addressed to more than one user, so that the total pairs (source, target) extracted from these messages was actually higher: 206.592. This is the number of directed arrows in our network. We stress again that our network is a dynamical instance of a larger underlying network (i.e., that made up of followers and followings in twitter). 

Finally, although not directly related to the study presented here but important for complementary studies, data for all the involved users were scrapped directly from {\em twitter.com} using a cloud of 128 different nodes of a subnet. The scrap was successful for 84.229 users,  for whom we also obtained their official list of followers. Moreover, about a half of them can be associated to a location (city), which is later translated to geographical coordinates via a standard geo-localization service from {\em Yahoo}. It is worth remarking that the extraction of followers gave a list in the order of 3 millions users, which roughly coincides with the order of the audience estimated by {\em Twitter} in Spain.

\section*{Results and Discussion}
The availability of time-stamped data allows us to closely track the birth and development of the network made up of users who exchanged tweets related to the 15M movement during the period analyzed. In this network, every node represents a user while a link between two nodes is established whenever they exchange a message. We have made a movie (see the {\em Supplementary Information} for a low resolution version or go to {\sl http://15m.bifi.es/index.php} for a high resolution one with downloading options) that reproduces the temporal evolution of the networks and the dynamics of messages exchange during the period analyzed. The reader will note the highly dynamic character of the network as well as the dimension of the social phenomenon being analyzed.

The network constructed as described above is weighted and directed, i.e., a link from $i$ to $j$ means that $i$ sent at least a message to $j$ and the weight of the link $i\rightarrow j$ stands for the actual number of such messages. Therefore, the adjacency matrix of the network is not symmetric ($j$ does not necessarily send a message to $i$). Moreover, it is worth noticing that we always work with accumulated data, such that the network at a certain time $t$ includes every message (link) produced at any time $t' \le t$, i.e., once a link is established, it connects the two end nodes forever. Once the network is built, we are able to characterize it from a topological point of view, within the framework of complex network theory \cite{boccaletti06,dgm08}. In doing so, we discuss local and global descriptors, as well as the network structure at the mesoscale level. Additionally, we also analyze the dynamics of information spreading over the 15M network and compare our results with those already reported for other dynamical processes that are mediated by the Web 2.0. 

\subsection*{Network Growth and Structure}
\begin{figure}[!ht]
\begin{center}
    \includegraphics[width=0.32\textwidth,clip=0]{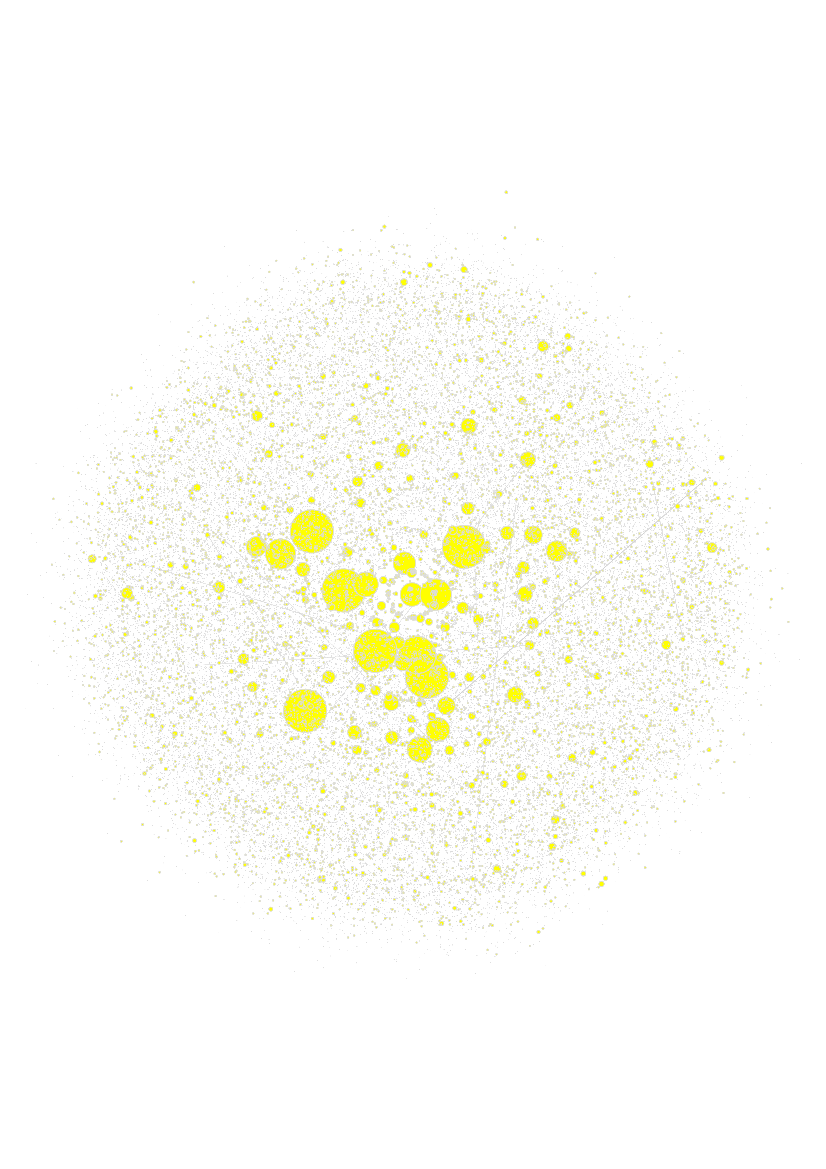}
    \includegraphics[width=0.32\textwidth,clip=0]{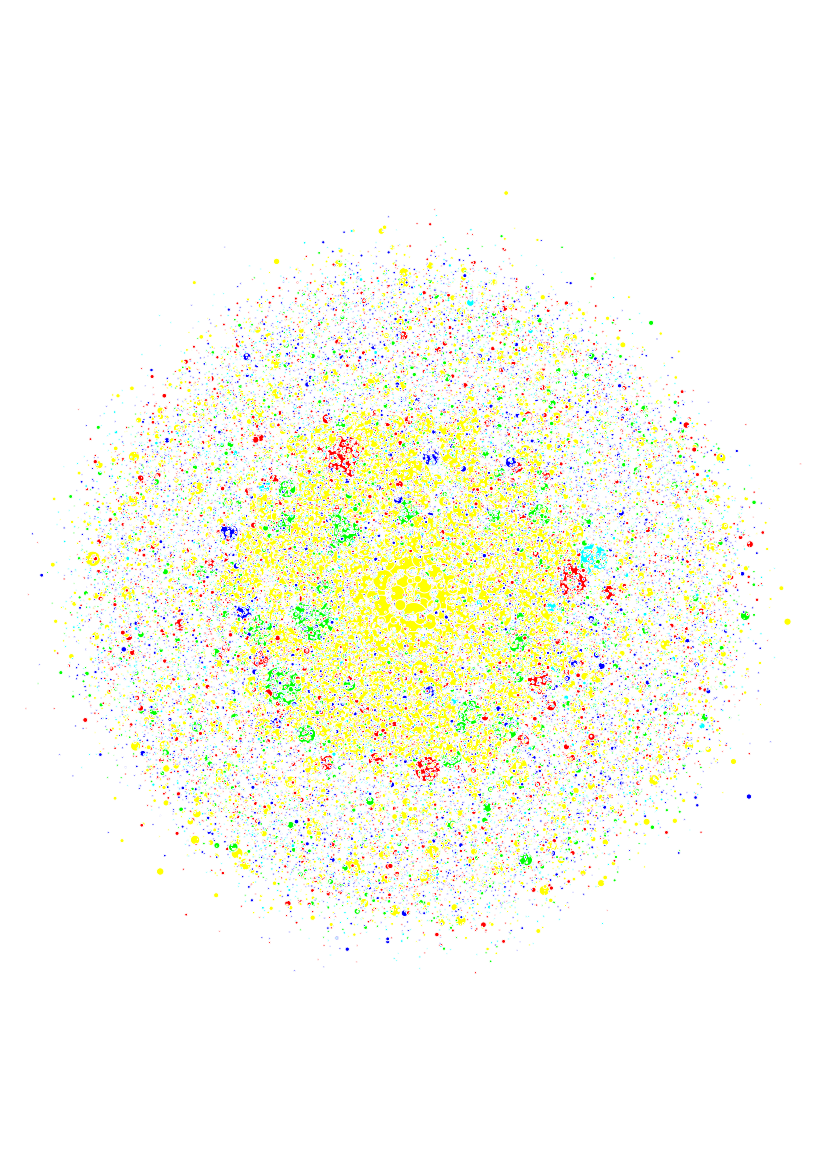}
    \includegraphics[width=0.32\textwidth,clip=0]{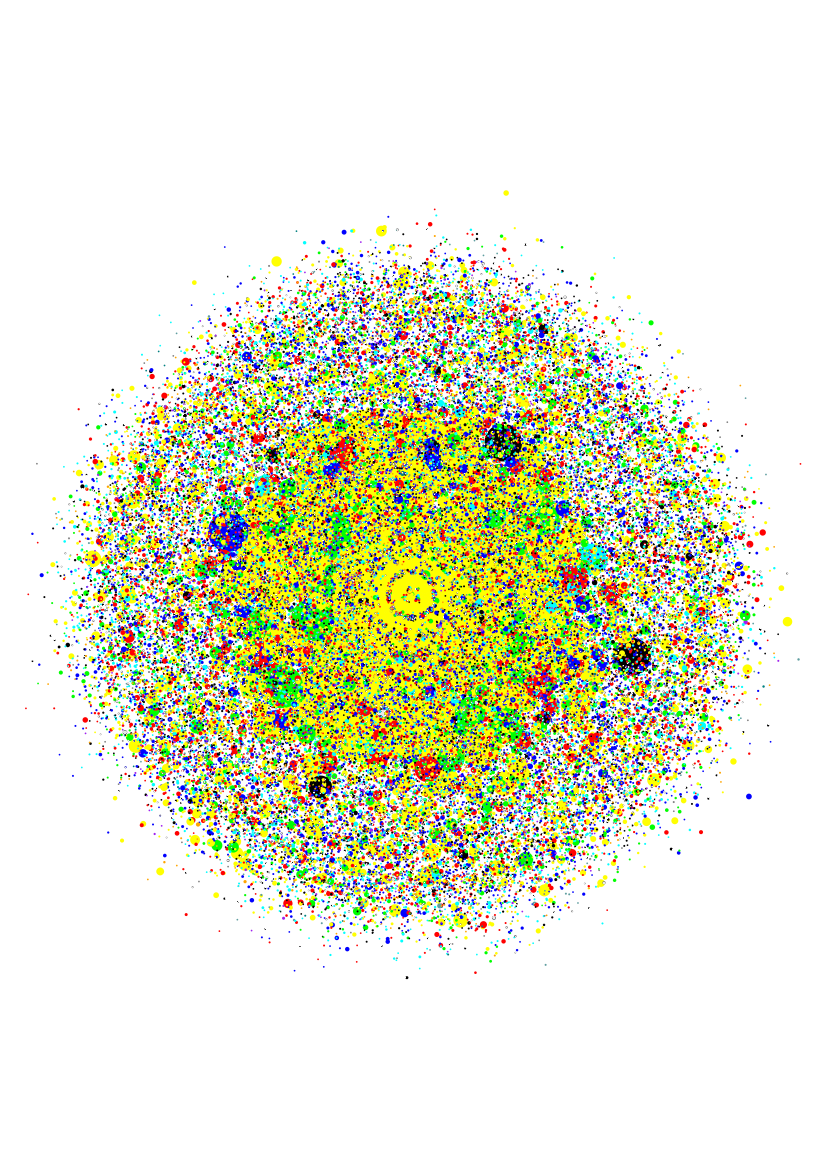}
    \includegraphics[width=0.8\textwidth]{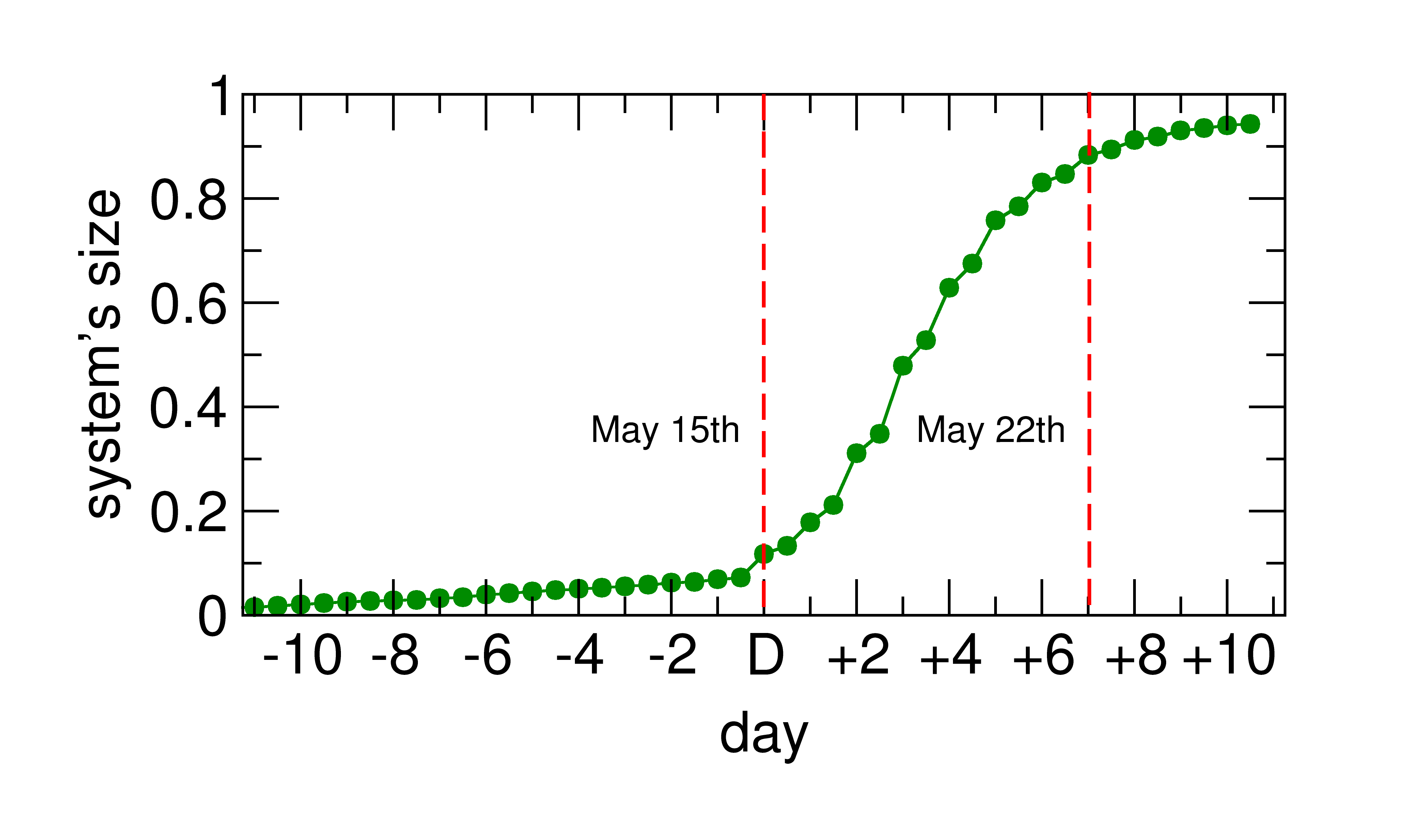}
\end{center}
\caption{{\bf Evolution of the 15M Network.} (Top) Cumulative snapshots of the system (at day $D$, $D+5$ and $D+10$). Different colors stand the age of the nodes (yellow for former adherents and black for latecomers). (Bottom) Evolution of the system's giant component, relative to the final size of the network. Note that growth does not proceed progressively, but it rather explodes from day $D$ due to socially relevant events on that date.}
\label{fig1}
\end{figure}

The first point of interest concerns the structural growth pattern. We wonder whether a collective mobilization of thousands of agents demands a slow, progressive increase in size; or rather social networking platforms enable an abrupt emergence. In Figure \ref{fig1} (top) we present three snapshots of the system for different days, relative to day $D$ (May 15th). Colors stand for the ``age'' of the node: early active users are coded in yellow, where those that adhere the network in successive days are coded in green, red, etc. Black is left for the latest adopters (people whose activity began at $D+10$). Besides, the size of the nodes has been made proportional to their activity, taking into account both incoming and outgoing tweets (however, for the sake of clarity, such proportion has been truncated at $k_{in}+k_{out}=200$ in the networks displayed in the figure). Even this simple representation of the evolution of the 15M network is already indicative of the growth in the number of agents once the movement took off and time goes on.

The results depicted in the bottom panel of Figure \ref{fig1} further illustrate the way in which the network evolves by gaining adherents. The figure represents the proportion of active nodes at time $t$ (with a resolution of 12 hours) in the giant component relative to the total number of users in the network at the end of the growth process. As we can see from the figure, the formation of the network and its later increase in size does not proceed in a gradual proportional process but in a sequence of bursts concentrated in just a few days (from day $D$ to day $D+7$). Obviously this process is driven by the events surrounding the movement: as mentioned, at day $D$ the protesters decided to camp at {\em Puerta del Sol} square in Madrid, which in turn elicited huge attention from the media and made the difference as far as the spread of the movement to a country wide scale concerns. Besides, from our data, it appears that the number of active users saturates after $D+7$. It is interesting to note that in May 21st ($D+6$), the day preceding local and regional elections, more than the 80\% of the network was already formed.  

Beyond structural growth, a second key aspect of the network under study concerns the distributions of strengths. The strength $s$ of a given node $i$ is defined, as usual, by the sum of the weights of the links that are incoming and outgoing to node $i$. In our case, it is also important for the discussion that will follow, to further divide this magnitude into two contributions. One the one hand, we have the strength derived from the weights of links incident to the node, $s_{in}$. This magnitude accounts for the total traffic (number of tweets) that a given node receives from its neighbors. Conversely, $s_{out}$ represents the sum of the traffic generated at a node, i.e., the number of tweets this user sends out. Additionally, let $P(s_{in})$ and $P(s_{out})$ be the cumulative distributions of both strengths, which we can be measured at different instants $t$ of the network development. 

\begin{figure}[!ht]
\begin{center}
  \includegraphics[width=0.8\textwidth]{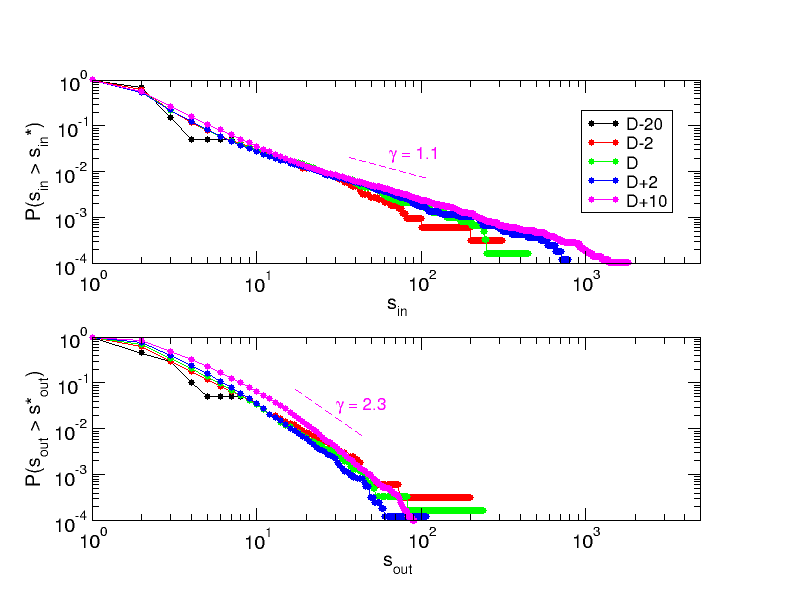}
\end{center}
\caption{{\bf Cumulative in- and out-strength distributions.} Strength distributions for both received (top) and sent (bottom) messages display a power-law behavior as early as at $D-2$. The fat-tailed distributions indicate that the 15M network is scale-free, with the implications this fact bears. Note that the exponents that define the power-laws differ significantly between sent and received messages.}
\label{fig2}
\end{figure}

Figure \ref{fig2} shows the cumulative distributions of the previous quantities for several times. As can be seen, even before the occurrence of the events that triggered public protests on day $D$, both $P(s_{in})$ and $P(s_{out})$ follow power-laws $P(s)\sim s^{-\gamma}$, but with different exponents ($\gamma_{in}=1.1$ and $\gamma_{out}=2.3$, respectively, as measured at $D+10$). Similar plots for the degree of the nodes exhibit the same behavior. It is well-known that the statistical properties of these variables in other technological, social and natural systems are also heterogeneously distributed. Therefore, the fat-tailed distributions that characterize the topology of the 15M network are not unique, but are rather ubiquitous in Nature. Nonetheless, the fact that the 15M network is scale-free, has deep consequences regarding a number of relevant issues including its origin, complexity, robustness and, from a dynamical point of view, the way in which information flows over the system. As the network obtained comes from the activity of the nodes, the heavy-tailed distribution of both nodes' degrees and strengths suggests a dynamics lacking any typical or characteristic scale. 

On the other hand, the dynamical asymmetry between incoming and outgoing degrees or strengths is not surprising neither. Indeed, individual behavior, which ultimately determines the resulting (out) dynamics, is an intended social action, but the emergent properties of the collective behavior of agents are unintended \cite{rybski2009scaling}. Essentially, subjects decide when and to whom a given message is sent. Therefore, the aggregate behavior of all agents and their popularity (i.e., how many incoming links a node has) result from individual choices. This is what is reflected in the in and out distributions. As a matter of fact, the exponent of the power law characterizing the degree probability distribution $p(k)$ lies in the interval $(2,3]$, as usually found in most real-world networks. Interestingly, spreading dynamics such as rumor and disease propagation processes are most efficient for scale-free networks whose exponent is precisely in this range \cite{vespiromu,mpsv02,rumor}. Finally, the strength distribution for the tweets sent, $p(s_{out})$, also resembles a power law function with an exponent larger than 3, although in this case the distribution exhibits an exponential cut-off. This might be due to the fact that sending messages has an associated cost in terms of bandwidth availability, the cognitive capacity to produce different messages and ultimately an unavoidable physical limitation to type them \cite{dunbar1,dunbar2,goncalves}.

\begin{figure}[!ht]
\begin{center}
  \includegraphics[width=1.0\textwidth]{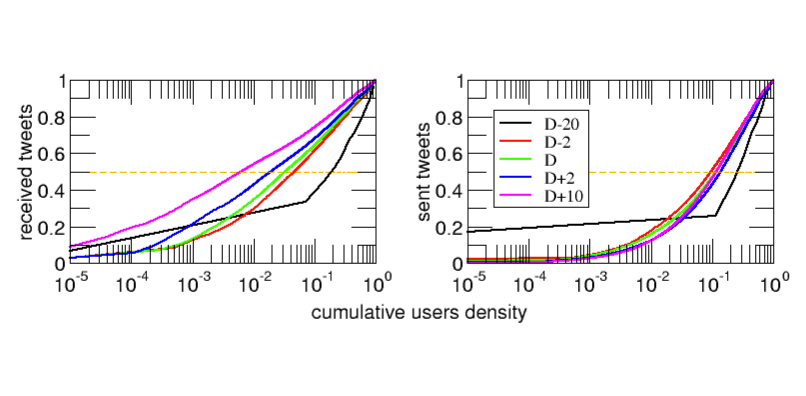}
\end{center}
\caption{{\bf Information flow.} The figure represents the density of tweets received (left) and sent (right) as a function of the cumulative fraction of active users. For each day, data are normalized by the number of active users at that date. As a reference, the horizontal line corresponds to 50\% of emitted/received tweets. Note that, on $D+10$, less than 1\% of the nodes receive half of the messages. On the contrary, the pattern of tweets sent hardly evolves from the beginning of the movement: 10\% of the active nodes produce 50\% of the messages. This asymmetry is coherent with the differences observed for the strength distributions.}
\label{infoctrl}
\end{figure}

Another aspect of capital interest regards how the overall traffic is generated and delivered. One of the main consequences of the functional form of the strength distributions is presented in Figure \ref{infoctrl}. The emergence of hubs, namely, the signaling feature of scale-free networks, leads to a predictable oligopoly in the way information is spread. In Figure \ref{infoctrl}, we observe that the number of tweets sent grows with the number of active users of the network. The curves corresponding to different days (i.e., instances of the network) nearly collapse into a single one. This means that as users join the network, the traffic generated scales accordingly. Moreover, the figure indicates that, for instance, roughly the 10\% of active subjects generate the 52\% of the total traffic. This is another indication of the dynamical robustness of the network to random failures but at the same time of its fragility to attacks directed towards that 10\% of users. More remarkably, the results depicted in the figure are in sharp contrast with the activity patterns corresponding to received tweets. In this latter case, as time goes on, the number of in-strength hubs decreases. As shown in the figure, by $D+10$, less than 1\% of users receive more than 50\% of the information. As we will show later on, these nodes correspond to authorities or mass media, which the adherents identify as main receptors (government) of or potential spreaders (mass media) for their messages. However, what at a priori seems to be a good choice, turns out to be harmful for the process of information spreading. As a matter of fact, we have checked that these hubs, which we call {\em information sinks} do receive a lot of messages but rarely act as spreaders within the network. As a consequence, almost all messages that arrive to those nodes are not redelivered and hence lost. In this sense, our results show that while the delivering of information is shared by a relative large number of users that keep the "social temperature" of the movement, most of this information is simply directed towards a few highly connected targets that might not pass the voice any longer (i.e., they are not active spreaders). Nonetheless, the information exchanged is public and users can therefore access it. This would however imply an individual action (to check a given user's timeline) that is not captured in our twitter data.

\subsection*{Community Structure}

The modular structure is pervasive in many natural, social and technological networks. Generally speaking, modules are islands of highly connected nodes separated by a relatively small number of links. This meso-level skeleton is likely to be relevant to understanding dynamical processes in networked systems. Agents in social networks tend to gather with those who share cultural traits (homophily) or professional interests \cite{arenas04,lozano2008community,blondel2008fast,fortunato2010community}, and more specifically, political communication networks tend to exhibit a clustered structure along political opinion lines \cite{adamic2005political,conover2011}.

\begin{figure}[!ht]
\begin{center}
  \includegraphics[width=0.8\textwidth]{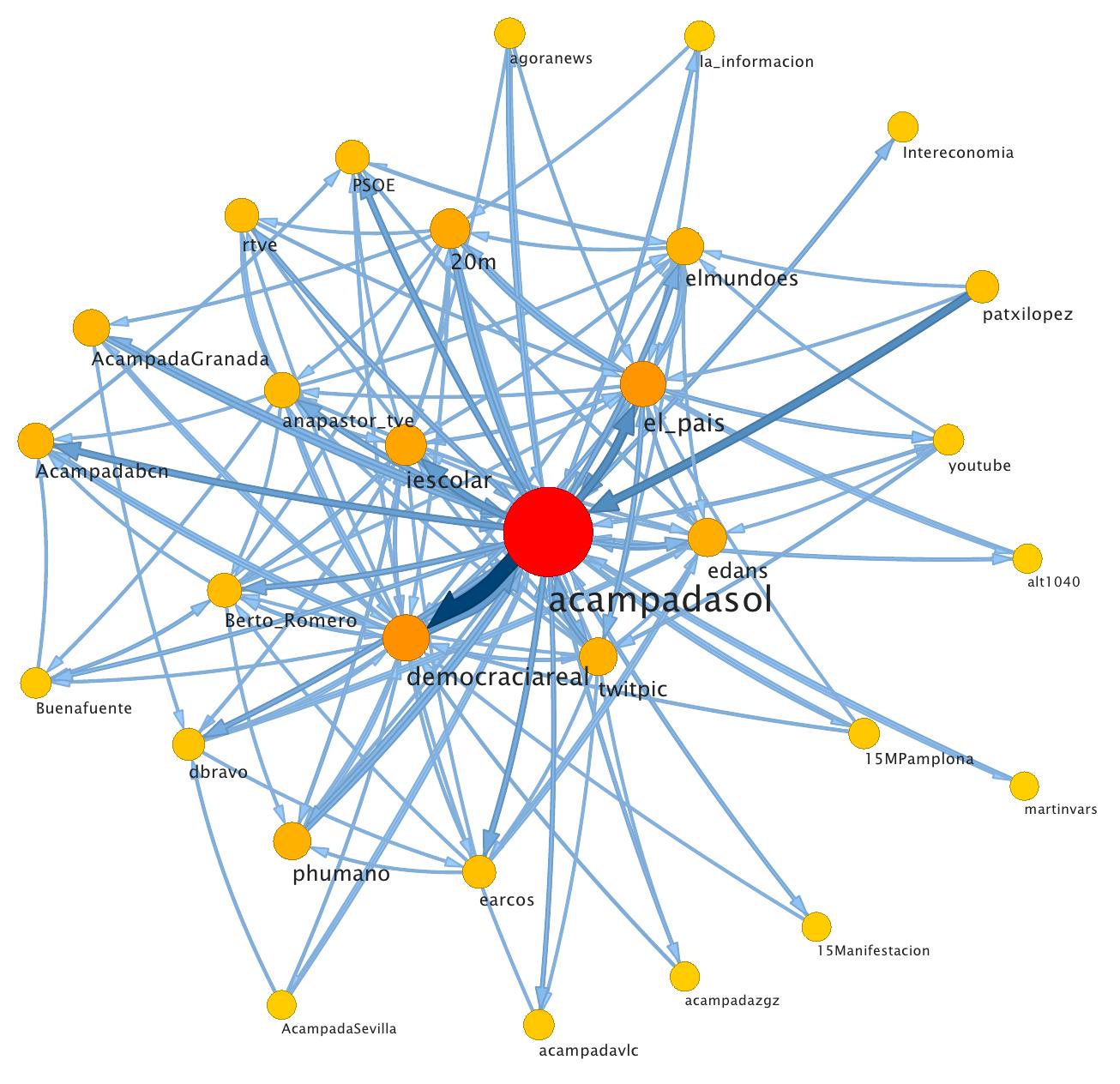}
\end{center}
\caption{{\bf Community structure of the 15M network} The figure shows, in a compact view in which each node represents a community, the 30 most important modules. They can be identified by a single node, around which the community is organized. These {\em local hubs} (labeled in the figure) agglutinate modules and act as information bridges connecting the whole network.}
\label{30coms}
\end{figure}

We have analyzed the community structure of the 15M network once its size stabilizes, i.e., at $t=D+10$. We have applied a random walk-based algorithm that optimizes a map equation on a network structure \cite{rosvall08}. Although alternative community detection algorithms are at hand, we chose the previous strategy because it is suited for networks, as it is actually our case, in which the dynamics of information flow is relevant. These type of algorithms rely on the intuitive idea that, if communities exist, a random walker tends to get trapped in them due to their dense within-connectivity \cite{pons05,rosvall08,borge10epjb}. The output of such information theoretic algorithm is a partition made up of 6388 modules. Most of these communities have less than ten nodes. We focus our analysis on the 30 most important modules from a dynamical perspective, i.e. those which concentrate most of the random walker's activity. These modules do not necessarily coincide with the first 30 communities ranked according to their size, but all of them contain over 100 nodes. Figure \ref{30coms} shows these 30 communities in a compact view (each node represents a community) \cite{edler2010mapequation}. Furthermore, each community is assigned a tag, corresponding to the most central node in that community. Again, these nodes have been identified as being dynamically dominant within their modules, thus they play an outstanding role in the dynamics of information. Our results show that modules are highly hierarchical and that nodes that are central to their communities, i.e., {\em local hubs}, are mostly hubs at the global scale as well.

The mesoscale structure allows to get deeper insights into social aspects of our case study. First, tags identifying the 30 largest communities are highly heterogeneous. 6 of these modules correspond to important mass media (newspapers and television), which points to the otherwise intuitive fact that users rely on these agents to amplify their opinion. The same can be said of 3 modules corresponding to famous journalists. More interestingly, 7 modules correspond to on-line activists and/or veteran bloggers. These agents are unknown to most people, but they are present in the network from its birth and enjoy a solid reputation that facilitates their being considered a reference in the movement. Remarkably, 7 modules are formed by camps in 7 different cities. Madrid is of course the main one, as the movement began there ({\em acampadasol}, which comprehends over 3000 nodes). Other cities are Barcelona, Granada, Zaragoza, Valencia, Seville and Pamplona. 

The fact that communities are geographically defined suggests some additional conclusions: (i) the mesoscale reflects the autonomy of each of the assemblies throughout the Spanish geography. Each of these modules hardly connects to any other, indicating a low communication between them; (ii) the exception to the previous point is Madrid: each minor camp holds a strong communication interchange with the community represented by {\em acampadasol}. Taking points (i) and (ii) together, it can be safely said that the movement is highly centralized, because in most cases a peripheral settlement is only influenced by Madrid and one or two minor ones. Finally, (iii) despite the potential of Web 2.0 communication platforms, data indicates that these media are mostly used to communicate with geographically close people. In other words, the network is {\em global}, but communication is mainly {\em local}. This is further verified in Table \ref{geopos}, where we have summarized the percentage of people whose geolocated information coincides with that of the module (city). 
 
\subsection*{Popularity Evolution}
The Web 2.0 has brought to network science the challenge to deal with highly dynamic, changing structures. Besides source and target nodes, and a (perhaps weighted) link between them, one must now consider a new ingredient: time. In this context, an interesting issue is related to the evolution of particular nodes: understanding how an element (be it a Wikipedia entry or a novel trend in social networks) comes to existence (appears in the network) and develops. Of further interest is to elucidate how a subset of these network's components ends up as a ``popular'' entity. This is a key aspect in network development, as popular agents eventually have an impact on other agents' opinions, acting as a referent, be those opinions related to politics, culture or business.

\begin{figure}[!ht]
\begin{center}
  \includegraphics[width=1.0\textwidth]{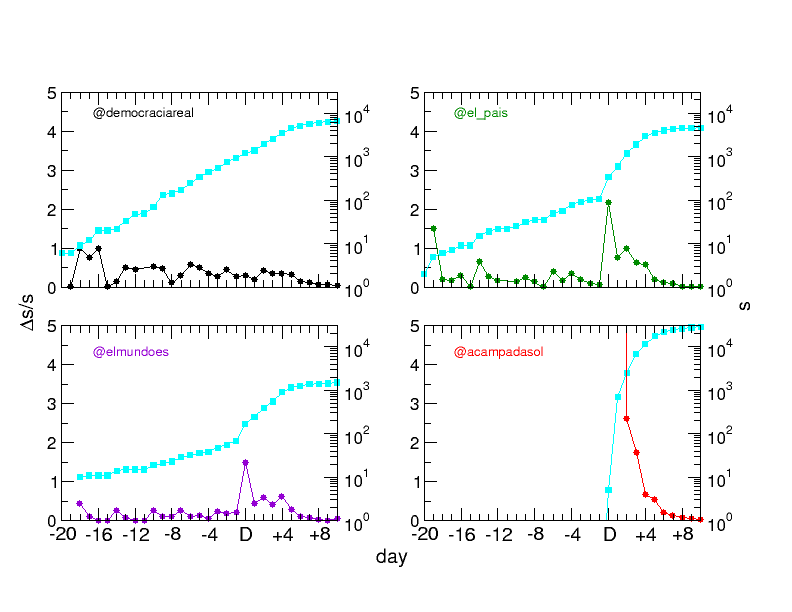}
\end{center}
\caption{{\bf Popularity dynamics.} The increase in the strength of the selected nodes (cyan curves) markedly changes after day $D$. The logarithmic derivative $\Delta s/s$ provides a finer interpretation in terms of the bursts in popularity. {\em democraciareal} (upper-left) had been an active agent long time before $D$, seemingly the movement was gathering strength (note some remarkable increments between $D-20$ and $D-16$). A newer user in the social network ({\em acampadasol}; lower-right) emerges and quickly takes over from $D$ onwards as the reference of the whole movement. On the other hand, two right- and left-wing newspapers, {\em El Mundo} and {\em El Pa\'is} respectively, undertake similar changes, which indicates that a large sets of nodes identified them as relevant actants in the context of protests. Note, however, that this does not imply that the same nodes are active senders too. As a matter of fact, they are not.}
\label{dknodes}
\end{figure}

To capture the dynamics of popularity, we follow the framework recently proposed in \cite{rybski2009scaling,ratkiewicz2010characterizing}. The natural quantity to measure popularity in a communication network is the number of messages that arrive at a node, which corresponds to that node's in-strength $s_{in}$, and the rate at which $s_{in}$ changes. Hence, a way to grasp how the activity of a node evolves is to consider its logarithmic derivative $[\Delta s/s]_{t} = (s_{t}-s_{t-1}) / s_{t-1}$, i.e., the relative variation of strength in a time unit (we omit subindex ``in'' for clarity). Figure \ref{dknodes} displays the evolution of the latter variable for some arbitrarily chosen nodes among those that are information sinks. Beyond an initial surge typically observed in many nodes, the time series of the logarithmic derivative evidence a bursty behavior. Fluctuations depend on exogenous events, in a strong parallelism with the external circumstances that drive the whole network's strong changes. It is noteworthy that these patterns closely resemble other, less conflictive, examples of popularity evolution in the Wikipedia or the Web \cite{ratkiewicz2010characterizing}. 

\begin{figure}[!ht]
\begin{center}
  \includegraphics[width=0.8\textwidth]{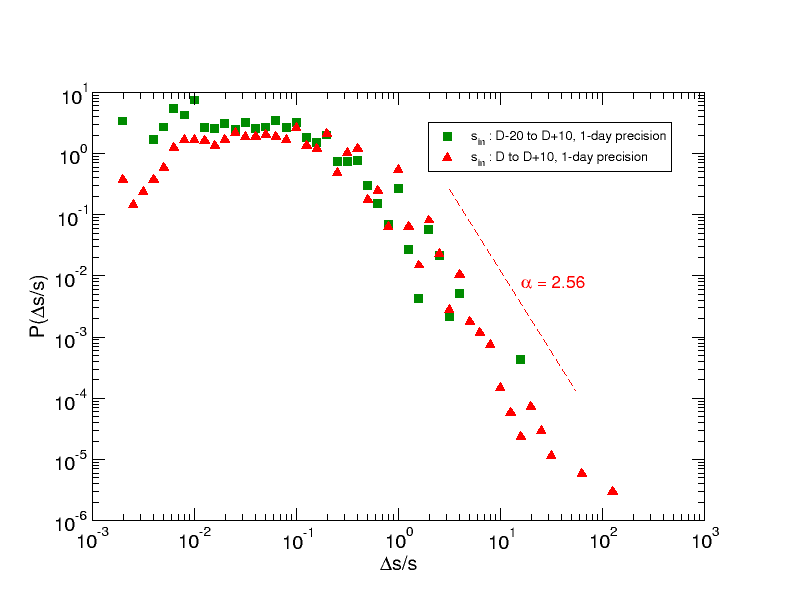}
\end{center}
\caption{{\bf Burst size distribution.} Popularity ``efervescence'' in the whole system (each node). Unsurprisingly, most nodes hardly undergo noticeable changes in their popularity. However, a small fraction of nodes does experience significant increases (heavy tail). This pattern is not exclusive of this communication network.}
\label{dkdistrib}
\end{figure}

\begin{figure}[!ht]
\begin{center}
 \includegraphics[width=0.8\textwidth]{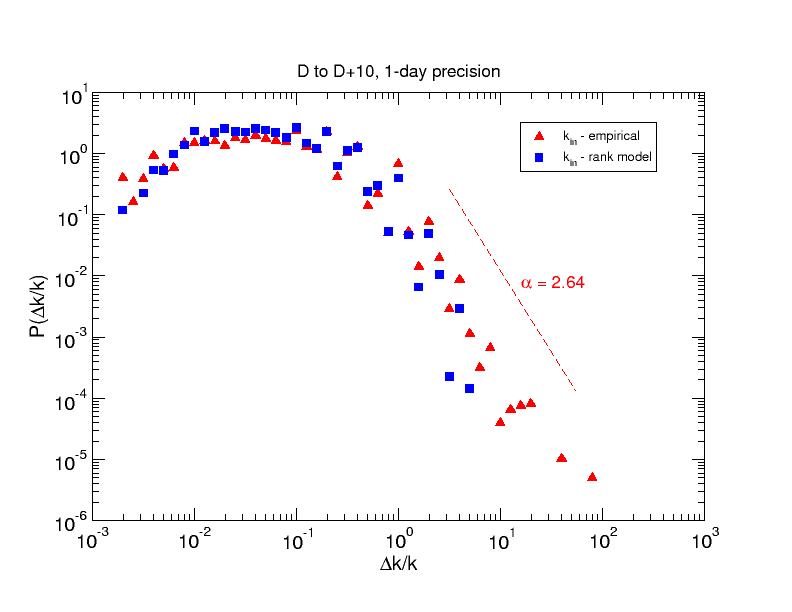}
\end{center}
\caption{{\bf Empirical and simulated burst size distribution.} The
rank model, even at its simplest form, successfully reproduces not
only empirical structural features, but also the prestige or
popularity evolution over time. See the text for more details about the model implemented.}
\label{rmodel}
\end{figure}

On the other hand, Figure \ref{dkdistrib} shows how bursts are distributed according to their magitude for two different time intervals but the same time granularity (1 day). The observed pattern, which is the same regardless of the time intervals under consideration, clearly shows heavy-tailed distributions, again in close resemblance to results already reported for other web-mediated dynamics \cite{ratkiewicz2010characterizing} and a variety of critical phenomena in physical, economic and social systems. As a matter of fact, a simple model can account for the observed bursts distribution. The so-called {\em rank model} is specially conceived for networks in which
prestige, rather than degree-based preferential attachment, plays a central role to determine nodes' connectivity \cite{fortunato2006scale}. The rank model depends on a prestige measure that is used to rank nodes. In this model, the probability that a new node that joins the network at $t+1$ connects to an older one $j$ is given by
\begin{equation}
p(t+1 \rightarrow j) = \frac{R_{j}^{-\alpha}}{\sum_{i=1}^{t} R_{i}^{-\alpha}}
\end{equation}
where $R_{j}$ is the rank of node $j$ and $\alpha > 0$ determines the exponent $\gamma$ of the resulting power-law degree distribution $p(k)$, such that $\gamma = 1 + \frac{1}{\alpha}$. Figure \ref{rmodel} compares the bursts distributions resulting form the data and from the model. The results shown correspond to the case in which nodes are ranked according to their age and $\alpha$ has been set to $0.9091$ (for this value we obtained the best fit to the actual degree distribution of the empirical network). Moreover, we note that in order to simulate the non steady growth of the network, we have considered for the synthetic case that 1 day has gone by when the size of the network being generated coincides with that of the empirically assembled network for that time. 

As we can see from the figure, even under the simplest rank rule, the burst magnitude distribution is nicely reproduced. As for our specific social context, the previous results do not imply that the driving mechanism behind the evolution of the 15M network is that simple, but illustrate that bursty activity of this sort can be produced by generic mechanisms. Given that this self-organized activity is widespread in nature, there are no reasons to consider that the network has its origin in external actions. Nonetheless, it must be pointed out that although the rank model grasps the main observed trends, both regarding structural characteristics and dynamical facts, yet other important aspects extracted from the data are not captured by the model. These limitations demand model refinements beyond the scope of the present study. 

\section*{Conclusions}

The social phenomenon here presented as a case study is a collective endeavor that is expressed at many levels, ranging from public demonstrations and camps to the presence of news in the mass media. In this work we have analyzed data from time-stamped, online activity in a specific social networking site during the formation and stabilization of this social movement. From a scientific point of view, these data represent a challenge as most network studies typically deal with a static structure. 

Undoubtedly, there are many facts that can be easily identified as the grounds of the 15M movement. Among them, the world-wide economic crisis and the impact it has had on society. Nonetheless, the particular events that triggered the growth of the whole movement remain unknown and are beyond the scope of this work. Addressing these questions would probably require an in-depth semantic analysis of the contents of interchanged messages. From its onset, our statistical characterization of the communication network built from tweets exchanged between adherents (and opponents) reveals a strong resemblance to well-known phenomena in natural and manmade systems, which are admittedly self-organized. 

Additionally, the 15M movement also raises relevant questions with sociological consequences. We argue that information centralization (Figure \ref{infoctrl}), as well as patterns of popularity growth (Figures \ref{dknodes} and \ref{dkdistrib}) are indicative of a tendency towards a hierarchical structure. Opinion leaders emerge spontaneously and minor actants devote much energy to communicate with them (be it to have their ideas echoed, or to influence such leaders). This proclivity is coherent with economy of attention \cite{simon71}, i.e., the system tends to avoid the overabundance of opinions to prevent scarcity of attention, but raises doubts about the possibility of converging to an egalitarian social system in which information flows and is received in an efficient manner. As far as our analysis concerns, we have shown that in a dynamics such that the one at work for our case, a relative large number of information sources exist, which is behind the robust functioning of the system. Conversely, communication sinks, where information is lost, are also generated.

On the other hand, our analysis of the community structure reveals some interesting facts. Geo-centered modules are abundant, but ideological or fame-related ones are also remarkable. It is important to keep in mind that time-stamped data is the dynamic (or functional) result of the activity on top of a more stable underlying structure, that of ``following'' and ``followers'' in the social networking site. For this reason, the communication network we observe is ever-changing and we argue that modules within it have not a straightforward interpretation. 

Finally, we have studied the patterns by which nodes become increasingly more visible. Results indicate that popularity growth in the context of political conflict does not display significant differences from other, less fashionable examples. Popularity is dominated by a fluctuating behavior, and popularity burst distributions lack of a characteristic scale. This fact connects the dynamics of popularity with other critical phenomena in many natural and artificial systems.

In summary, online social networks and the Web 2.0 provide new challenges to network theory. Events in the real world, ranging from economic phenomena to political protest, stand as the driving forces leading to the emergence of complex, time-evolving communication patterns. In this scenario, network theory stands as a suitable tool to unfold the structural and dynamical facets of such emergent systems.

% You may title this section "Methods" or "Models". 
% "Models" is not a valid title for PLoS ONE authors. However, PLoS ONE
% authors may use "Analysis" 

% Do NOT remove this, even if you are not including acknowledgments
\section*{Acknowledgments}

We are indebted to Beatriz Antol\'i, Guillermo Losilla, Rub\'en Valles and Isabel Vidal for their help and assistance at several stages of this study. J.B.-H is supported by the Government of Arag\'{o}n (DGA) through a grant to FENOL and by Spanish MICINN through project FIS2008-01240. D.I. is supported by the Government of Arag\'on through a Fundaci\'on ARAID contract. Y. M. was partially supported by the FET-Open project DYNANETS (grant no. 233847) funded by the European Commission, by Spanish MICINN through projects FIS2008-01240 and FIS2009-13364-C02-01 and by Comunidad de Arag\'on (Spain) through the project FMI22/10. We also acknowledge the Spanish MICINN for financial support through the project FIS-164-50.

%\section*{References}
% The bibtex filename
%\bibliography{bibtesi}

\section*{Figure Legends}

\section*{Tables}

\begin{table}[!ht]
\caption{\bf{Geographic origin of nodes in region-based communities}}
\begin{center}
\begin{tabular}{|c|c|c|}
%table information
\hline
\bf{community tag}&\bf{area}&\bf{fraction of users from same area}\\\hline
@acampadasol&Madrid&54\%\\\hline
@acampadabcn&Barcelona&81\%\\\hline
@acampadavlc&Valencia&63\%\\\hline
@acampadazgz&Zaragoza&82\%\\\hline
@acampadagranada&Granada&53\%\\\hline
@acampadasevilla&Sevilla&83\%\\\hline
@15MPamplona&Pamplona&71\%\\\hline
\end{tabular}
\end{center}
\begin{flushleft} Region-based modules are mostly formed by nodes whose geographical origin coincides with that of the most central node in the community. This statement is clear for almost all these modules, except in the case of Madrid and Granada. The case of Madrid is not surprising, given that @acampadasol is the reference of the whole movement, thus the community organized around this actant is a more heterogeneous one. Granada is a more intriguing exception.
\end{flushleft}
\label{geopos}
 \end{table}

\end{document}